# Great Expectations:

# Plans and Predictions for New Horizons Encounter with Kuiper Belt Object 2014 MU$_{69}$ ('Ultima Thule')

Jeffrey M. Moore[1]*, William B. McKinnon[2], Dale P. Cruikshank[1], G. Randall Gladstone[3], John R. Spencer[4], S. Alan Stern[4], Harold A. Weaver[5], Kelsi N. Singer[4], Mark R. Showalter[6], William M. Grundy[7], Ross A. Beyer[1, 6], Oliver L. White[1, 6], Richard P. Binzel[8], Marc W. Buie[4], Bonnie J. Buratti[9], Andrew F. Cheng[5], Carly Howett[4], Cathy B. Olkin[4], Alex H. Parker[4], Simon B. Porter[4], Paul M. Schenk[10], Henry B. Throop[11], Anne J. Verbiscer[12], Leslie A. Young[4], Susan D. Benecchi[11], Veronica J. Bray[13], Carrie. L. Chavez[1, 6], Rajani D. Dhingra[14], Alan D. Howard[15], Tod R. Lauer[16], C. M. Lisse[5], Stuart J. Robbins[4], Kirby D. Runyon[5], Orkan M. Umurhan[1, 6]

[1]National Aeronautics and Space Administration (NASA) Ames Research Center, Space Science Division, Moffett Field, CA 94035, USA.
[2]Department of Earth and Planetary Sciences, Washington University, St. Louis, MO 63130, USA.
[3]Southwest Research Institute, San Antonio, TX 78238, USA.
[4]Southwest Research Institute, Boulder, CO 80302, USA.
[5]Johns Hopkins University Applied Physics Laboratory, Laurel, MD 20723, USA.
[6]The SETI Institute, Mountain View, CA 94043, USA.
[7]Lowell Observatory, Flagstaff, AZ 86001, USA.
[8]Massachusetts Institute of Technology, Cambridge, MA 02139, USA.
[9]NASA Jet Propulsion Laboratory, NASA Jet Propulsion Laboratory, California Institute of Technology, Pasadena CA 91109 USA.
[10]Lunar and Planetary Institute, Houston, TX 77058, USA.
[11] Planetary Science Institute, Tucson, AZ 85719, USA.
[12]Department of Astronomy, University of Virginia, Charlottesville, VA 22904, USA.
[13] Lunar and Planetary Laboratory, University of Arizona, Tucson, AZ 85721, USA
[14]Department of Physics, University of Idaho, Moscow, Idaho, 83843, USA.
[15]Department of Environmental Sciences, University of Virginia, Charlottesville, VA 22904, USA.
[16]National Optical Astronomy Observatory, Tucson, AZ 85726, USA.

**Main point #1**: New Horizons will fly to within 3500 km of 2014 MU69, acquiring images with pixel scale resolutions significantly better than 100 m/pixel.
**Main point #2**: Our spectroscopic observations potentially could detect $H_2O$, $CH_4$, $N_2$, $CH_3OH$, and $NH_3$, depending on their brightness and abundance.
**Main point #3**: We will evaluate Ultima's composition, surface geology, structure, near space environment, small moons, rings, and the search for activity.

**Index Term:** 6224 Kuiper Belt Objects

## Abstract

The *New Horizons* encounter with the cold classical Kuiper Belt object (KBO) 2014 MU$_{69}$ (informally named "Ultima Thule," hereafter Ultima) on 1 January 2019 will be the first time a spacecraft has ever closely observed one of the free-orbiting small denizens of the Kuiper Belt. Related to but not thought to have formed in the same region of the Solar System as the comets that been explored so far, it will also be the largest, most distant, and most primitive body yet visited by spacecraft. In this letter we begin with a brief overview of cold classical KBOs, of which Ultima is a prime example. We give a short preview of our encounter plans. We note what is currently known about Ultima from earth-based observations. We then review our expectations and capabilities to evaluate Ultima's composition, surface geology, structure, near space environment, small moons, rings, and the search for activity.

## 1. 2014 MU$_{69}$ in the Context of the Cold Classical Kuiper Belt

The cold classical Kuiper Belt object (CCKBO) 2014 MU$_{69}$ ('Ultima') is the flyby primary target for NASA's *New Horizons* first Kuiper Belt Extended Mission (Stern et al., 2018). The cold classical Kuiper Belt consists of objects on non-resonant, low-eccentricity (*e*), low-inclination orbits (typically $i < 5°$ to the invariable plane of the Solar System), i.e., dynamically cold (relatively unexcited) orbits, with heliocentric semimajor axes (*a*) between about 40 and 50 AU. The cold classical objects likely formed in-place and escaped perturbation from their initial orbits by giant planet migration (Parker & Kavelaars, 2010; Batygin et al., 2011; Dawson and Murray-Clay, 2012, Fraser et al., 2017), making them the most distant known remnants of the original protoplanetary disk. Ultima's orbital elements ($a = 44.2$ AU, $i = 2.4°$, $e = 0.03$) (Porter et al., 2018) are fully consistent with it being a CCKBO, as is its visible color (discussed below).

The cold classical KBOs are on average physically smaller, that is, lacking in large bodies (dwarf planets), as well as following a steep power-law size-frequency distribution at the bright, or large end (Petit et al., 2011; Fraser et al., 2014), and they are spectrally redder in the visible and near-infrared than the dynamically hot KBO populations (e.g., Tegler & Romanishin, 2000; Doressoundiram et al., 2008). The dynamically hot KBO populations (hot classicals, bodies in mean-motion resonances with Neptune such as Pluto, scattered disk bodies, etc.) are thought to have formed closer in to the Sun, and have been scattered and emplaced in their present orbital configurations by giant planet migration (principally, that of Neptune) (Nesvorný & Vokrouhlický, 2016, and references therein). A much larger fraction of the cold classicals are wide binaries as well (at least 30%), compared with the hot classicals (Noll et al., 2008; Parker et al., 2011; Fraser et al., 2017). The steep size distribution and preponderance of binaries suggest that the cold classical population has never been subjected to a period of intense collisional comminution (Nesvorný et al., 2011; Parker & Kavelaars, 2012), unlike other KB sub-populations. The cold classicals thus appear to be a physically and dynamically distinct, and likely a more pristinely preserved population.

The upcoming *New Horizons* flyby of Ultima on 1 January 2019 is a unique opportunity to explore the disk processes and original chemistry of the primordial solar nebula, as well as the nature of CCKBO bodies themselves. Naturally, compositional measurements during the flyby are of paramount importance, as is high-resolution imaging of shape and structure, because the intermediate size of Ultima (much smaller than Pluto but $\sim 10^3 \times$ more massive than a typical comet) may show signs of its accretion from much smaller bodies [layers, aggregates, lobes, etc., in the manner of comet 67P/Churyumov-Gerasimenko (Massironi et al., 2015; Rickman et al., 2015; Davidsson et al., 2016; Jutzi & Benz, 2017; Nesvorný et al., 2018)], or alternatively, derivation via the collisional fragmentation of a larger body if KBOs are "born big" (cf. Morbidelli & Rickman, 2015; Nesvorný et al., 2010; Johansen et al., 2015; Schwartz et al., 2018). Ultima is also large enough (of order 20 km $\times$ 35 km and possibly a binary; Buie et al., 2018) that it may show some signs of internal evolution driven by radiogenic heat from $^{26}$Al decay, if it accreted early enough and fast enough (Prialnik et al., 2007; McKinnon et al., 2008).

## 2. Highlights of the Planned Observations by *New Horizons*

A detailed account of this subject is given in Stern et al. (2018); here we briefly summarize: The major science goals of the *New Horizons* encounter with Ultima are to: (1) characterize its global geology, morphology and rotational characteristics; (2) map its surface composition; (3) search for any satellites and rings and study them as feasible; (4) characterize or constrain its composition and magnitude of any volatile or dust escape from Ultima; (5) characterize the physical properties of its surface; (6) determine its crater size/frequency distribution; (7) constrain its bulk parameters (mass, density); and (8) evaluate evidence of interaction with solar wind. At closest approach *New Horizons* should pass Ultima at a nominal distance of 3500 km and at a relative velocity of 14.16 km/sec.

On board *New Horizons* is a suite of remote sensing imagers and fields and particles detectors (see Weaver et al., 2008; for overview). The instruments most relevant to geological and compositional investigations are the LORRI, LEISA, and MVIC, the specifications of which are described in detail elsewhere (Cheng et al., 2008; Reuter et al., 2008; Stern et al., 2018). Because the rotational period, shape, and rotation axis orientation of Ultima are not known, it is unclear how much of the object will be seen at high resolution. As described in more detail in Stern et al. (2018), resolved observations of Ultima will begin about 2.5-3 days from the closest approach time (~5:33 UTC). Starting at 39 days to closest approach, unresolved images taken by LORRI will be used to search for rings and satellites, especially in regards to potential navigational hazards to the spacecraft. For ~22 hours preceding the encounter an alternating series of LORRI, MVIC, and LEISA observations will be taken with an increasing cadence designed to be robust to observing different longitudes on the object given different rotation periods. For example, at ~12 hours out, the cadence of observations is every 60 minutes, while at ~3.5 hours out the cadence is tightened to observations every 20 minutes. During the 24 hours before closest approach there will also be satellite and ring searches of a broader region of space around Ultima.

Beginning approximately 80 minutes before closest approach, a series of LORRI images will be taken as ride-along observations during LEISA and MVIC scans of the nominal position of Ultima and the downtrack error ellipse around Ultima, derived from the uncertainty in Ultima's orbit. These observations will provide the highest resolution panchromatic, color, and spectral data on Ultima (Figure 1). LORRI images may be as good as ~35 m/px (with some degradation from smear in one direction) but will generally have lower signal-to-noise than the panchromatic MVIC observations, which reach pixel scales as high as ~135 m/px (and should have little-to-no smear). The nominal phase angle of ~32° for the highest resolution observations should allow for reasonable visual analysis of topographic features (Figure 2).

Stereo observations are constrained by the close flyby nature of the encounter. Optimal stereo parallax angles are 20 to 30°. The best convergence angle under the nominal flyby plan is approximately 21° (combining the closest approach image at 32° phase and images from the incoming trajectory at 11° phase), but these images also have—at best—a factor of four difference in ground scale (which will make feature correlation difficult). There are observations that provide more similar ground scales, but at the cost of poorer parallax angles (~15°). The best terrain model we expect to be able to make will be approximately 1 km/pixel in ground scale and will have a stereo height accuracy around 300 m. These estimates are based on the assumption that Ultima will not have rotated much in the 20 minutes to an hour between these observations. If Ultima is a fast rotator, such that these planned observations do not capture the same portion of the surface, then stereo correlation will be significantly more challenging. Shape from shadows and shading will have to suffice.

## 3. Compositional Knowledge and Speculations

Ices detected on planetary satellites, Centaurs, Pluto, and Kuiper Belt objects in the outer Solar System (OSS) include $H_2O$, $CH_4$, $N_2$, CO, $CO_2$, $CH_3OH$ (methanol), HCN (hydrogen cyanide), $NH_3 \cdot nH_2O$ (ammonia hydrate), and $C_2H_6$ (ethane) (Clark et al., 2013). Most of these represent the inventory expected for low-temperature condensation in the outer parts of the solar nebula (e.g., Lewis, 1972), and a few can be synthesized by chemical processing of the others. At the low temperatures in the OSS the volatility of these ices ranges over many orders of magnitude from the most volatile ($N_2$ and CO) to the least ($H_2O$). Watson et al. (1963) and Lebofsky (1975) calculated the lifetimes against evaporation of icy bodies in the OSS, finding that $H_2O$, $CO_2$, $NH_3$ and $NH_3$ hydrates are highly stable for the age of the Solar System, while $CH_4$ evaporation is slow but not negligible. Methane is readily processed by ultraviolet light with charged particles forming a refractory colored solid, so the original $CH_4$ inventory of a small icy body such as Ultima Thule may be locked up in such a material. Ar, Xe, $O_2$, CO, and $N_2$, where exposed at the surface, are expected to have evaporated from Ultima Thule long ago.

The ability of the LEISA imaging spectrometer on *New Horizons* to detect ices on Ultima depends on the presence of suitably strong spectral absorption bands in its operational wavelength range of 1.25 - 2.5 µm and native spectral resolution $\lambda/\Delta\lambda = 240$ (Reuter et al. 2008). $H_2O$, $CH_4$, $CH_3OH$, and $NH_3$ meet these criteria; in data of high signal precision (signal/noise ratio), $NH_3$ hydrates can also be detected, as they were on Pluto's satellites Charon, Nix, and Hydra (Cook et al., 2018). The highest spatial resolution LEISA observation of Ultima will be ~1.8 km px$^{-1}$, and the highest resolution MVIC color observation will be ~0.3 km px$^{-1}$.

Ultima has an estimated albedo $p_V = 0.068 \pm 0.006$ calculated from $m_V \approx 27.5 \pm 0.1$ (Porter et al., 2018; Benecchi et al., 2017) and an equivalent radius of 15 km; occultation results suggest Ultima could be a binary or bi-lobate object of 20 x 35 km size (Buie et al., 2018), implying a somewhat higher $p_V = 0.087 \pm 0.008$. Unpublished measurements with the Hubble Space Telescope show that MU69 is red in color, a property shared with many small bodies in the outer Solar System (S. Benecchi, personal communication 2018; Benecchi et al., 2017). In contrast, Pluto's small satellites Nix and Hydra have much higher visible albedos (0.56 and 0.83, respectively) and generally neutral colors, although Nix has a large red-colored region that may be indicative of an organic component derived from radiation processing of a native hydrocarbon such as $CH_4$ (Weaver et al., 2016). The presumed original $CH_4$ is not, however, detected in the near-infrared spectrum of Nix, which instead is dominated by $H_2O$ ice absorption bands plus a band at 2.21 μm attributed to an ammonia hydrate (Cook et al., 2018).

Ultima Thule's low albedo may preclude the detection of any spectral bands attributable to ferro-magnesian silicates or ices, although bands of $H_2O$ ice could be observed; in models of mixtures of $H_2O$ with dark particles, the absorption bands of water ice at 1.55, 2.0, and 2.3 μm are visible. For example, Phobos has an albedo of 0.07 and no ice bands are seen, while the extremely red-colored Centaur (5145) Pholus has a similarly low albedo at 0.5 μm and clearly detected absorption bands of $H_2O$ (2.0 μm) and $CH_3OH$ (2.27 μm) ices (Cruikshank et al., 1998). The best-fitting model of the Pholus spectrum included magnesium-rich olivine (Fo82), and suggests that ferro-magnesian silicates might occur together with ices on Ultima Thule's surface. In the inner Solar System, nanophase neutral iron sputtered from Fe-bearing minerals and deposited on surface grains accounts at least in part for the red colors seen on the Moon and some asteroids, while larger Fe particles contribute to an overall darkening of their surfaces (Pieters & Noble 2016). The mechanism, efficiency, and long-term effects of this or a similar space weathering process at 44 AU are unknown. Saturn's irregular satellite Phoebe, thought to be a captured object from the Kuiper Belt (Johnson and Lunine, 2005), has a visible albedo $p_V = 0.0856 \pm 0.0023$ (Miller et al., 2011) and also has clearly detected absorption bands of $H_2O$ (Buratti et al., 2008). Localized exposures richer in ice may also exist. Ultima, however, is much closer in size to Phobos (mean radius 11.3 km) than either Phoebe (mean radius 106.5 km) or (5145) Pholus (mean radius 50 – 85 km).

As a summary prediction for the properties of Ultima detected with LEISA and MVIC, we might expect to find the $H_2O$ ice band at 2 μm, plus possible detections of $CO_2$, or $CH_3OH$ bands if they are sufficiently abundant, but probably no $CH_4$, and an overall reddish color suggestive of processed hydrocarbons. Alternatively, there may be red-colored regions superimposed on the otherwise more neutral-colored surface, suggestive of processed hydrocarbons exposed in the subsurface by impact(s), and/or fragmentation, perhaps indicating an irregular early history of accretion from an inhomogeneous source in the solar nebula. There may be outgassing pits or fissures indicative of an original component of $N_2$ or CO that has since been lost. In detail, Ultima is unlikely to look much like comet 67P/CG, which has significantly outgassed by passages close to the Sun and residence time in the inner Solar System (possibly losing ~400 m of surface material) (Nesvorný et al., 2018), thus affecting the texture, composition, and shape of its surface as revealed by *Rosetta*.

## 4. Inferences for Ultima's Gas and Particle Environment

As noted above, it is likely that any highly volatile species (e.g., Ar, Xe, $O_2$, $N_2$, CO, $CH_4$) that might have once been present on the surface of Ultima are now gone from its surface. However, it is possible that some less volatile ices (e.g., $CH_3OH$, $C_2H_2$, $C_2H_6$, HCN, $NH_3$) remain (e.g., Lebofsky, 1975; Schaller & Brown, 2007). Although these species are stable on the surface at their current temperature, ongoing radiation processing and occasional impacts provide a possible source for a transient atmosphere. Although unlikely, such a transient atmosphere will be searched for by the Alice ultraviolet spectrograph (Stern et al., 2008), e.g., (i) in absorption, (ii) using stellar and (iii) solar appulses, and (iv) in emission, using resonantly scattered solar emissions. Dust coma or material otherwise associated with Ultima will be searched for with high phase angle LORRI and MVIC imaging, and with *in situ* SDC observations. In addition, the particle and plasma environment of KBOs is largely unknown. *New Horizons* fields and particle instruments SWAP, and PEPSSI observations will establish the interaction of the interplanetary medium with Ultima, e.g., by looking for pickup ions resulting from sputtering of surface materials. Finally, *New Horizons* Alice, SWAP, and PEPSSI observations will characterize the fluxes of UV, solar wind, interstellar pickup ions, and energetic particles, i.e., space weathering, that modify KBO surfaces.

## 5. Geological Outlook and Speculations

The only other Kuiper Belt objects of comparable size to Ultima that have been imaged at pixel scales comparable to what *New Horizons* will achieve at Ultima are Pluto's small moons Nix and Hydra. Those moons are ~43 km and ~45 km in diameter, and were imaged by *New Horizons* at 0.3 km/pixel and 1.14 km/pixel, respectively (Weaver et al., 2016). *New Horizons* identified 11 craterlike features identified on Nix and 3 on Hydra. The rims of 3 km diameter craters on Nix have a softened appearance at 0.3 km/pixel, suggesting the possible presence of a mantling material (possibly impact ejecta) covering its surface (see below). Additionally, a faint linear fabric was observed on Nix, probably imparted by the surface expression of internal structure.

The presence of impact craters, a ubiquitous feature on most Solar System bodies, is almost certain on Ultima. However, Solar System objects that attain sizes comparable to that of Ultima (~30 km across) also have the potential to harbor unusual, and in some cases, possibly active, surface geological processes: several small satellites of Saturn, including Helene and Epimetheus, and Methone (in the extreme) display what appears to be fine-grained material covering large portions of their surfaces. These may be related to the dynamic ring- and mid-size moon region of the Saturn system of course (Umurhan et al., 2015). The surface of Phobos, the largest satellite of Mars, displays an unusual system of parallel grooves, as do many asteroids observed in this size range (Figure 2).

Although there have been no detailed predictions published of the collisional rate between Ultima and each of the Kuiper Belt sub-populations as was done for Pluto in advance of flyby (Greenstreet et al., 2015), we expect a cratering rate per $km^2$ of the same order as on Pluto and Charon (JeongAhn et al., 2018). Impact velocities should be somewhat lower than at Pluto, and extend down to the escape speed from Ultima (~10 m $s^{-1}$ for a mean density of 500-1000 kg $m^{-3}$). The low velocity tail could be more accretional than erosive, and such events may be efficient at producing ejecta that remains gravitationally bound, re-accreting onto the main body as a mantle.

Such a mantle can modify and remove existing craters, altering apparent crater density. Additionally, if a substantial mantle forms, it might become mobile and be transported across large portions of the body's surface by the seismic energy of later impacts.

Sublimation-modified landforms, commonly seen on comets and a number of other Solar System objects including Pluto, could have developed on Ultima even if it has never spent any time closer to the Sun than theory suggests for cold classical KBOs. The volatiles most able to sublimate and form erosional landforms would be surface exposures of $N_2$ and CO and other volatiles as delineated above. If exposed at the surface, $CH_4$ will also sublimate, though much more slowly than $N_2$ and CO. The loss of these volatiles, if present in sufficient quantities and concentrations and susceptible to exposure at the surface, might form distinctive and diagnostic scarps, pits and perhaps lags (e.g., Spencer, 1987; Moore et al., 1999). By corollary one can speculate that crater rim structures, if initially formed from deposits of volatile-rich material excavated from the interior, could show a unique range of pits, scarps, and lags surrounding the central depression of the original crater. Volatile loss may also mobilize grains in saltation or reptation, the principal grain motions in aeolian geology (e.g., Bridges et al., 2012), and these motions may manifest as large aeolian ripples as on 67P (Thomas et al., 2015; Jia et al., 2017). The scale of these ripples (low 10s of m wavelength) will be at the limit of the pixel scale of the LORRI camera, though location near the terminator would improve the chances of their detection.

Tectonic features might be reasonably anticipated on Ultima. Many small objects exhibit them, usually in the form of parallel scarps, troughs or coalescing strings of pits. On objects like the martian moon Phobos (Figure 2), these features are thought to be the consequence of martian tides and impact-induced seismic fracturing of the interior (e.g., Prockter et al., 2002; Buczkowski et al., 2008; Morrison et al., 2009). The appearance of such features on Ultima would provide insight into the interior strength and coherence.

Finally, the scattering properties of particles on Ultima's surface should become the prototype for those of other cold classical KBOs. From the disk-resolved analyses of spatially resolved *New Horizons* LORRI and MVIC images, the surface of Ultima will be the only CCKBO surface whose microphysical structure (macroscopic roughness, single particle phase function) has been characterized by disk-resolved photometric analyses. Furthermore, the disk-integrated solar phase curve derived from *New Horizons* observations of Ultima will provide context for the interpretation of the disk-integrated phase curves measured by *New Horizons* of other, distant cold classical KBOs (Verbiscer et al., 2017; Stern et al., 2018) and enable comparisons between the physical characteristics of particles on their surfaces with those on Ultima's. Comparing Ultima's disk-integrated phase curve with those of other objects in the OSS should reveal whether particles on its surface preferentially scatter incident visible sunlight in the forward direction like those on both Nix and Hydra (Verbiscer et al., 2018) and Phoebe (Buratti et al., 2008).

## 6. Shape and Structure

Results of the 17 July 2017 stellar occultation by Ultima indicate that this object displays a very irregular shape and could even be a close or contact binary (Buie et al., 2017, 2018; Zangari et al., 2017). Binaries are common in the Kuiper Belt (Noll et al., 2008), including contact binaries (Lacerda, 2011; Lacerda et al., 2014; Thirouin et al., 2017; Thirouin and Sheppard, 2017), and it has been argued that many short period comets, such as 67P/Churyumov–Gerasimenko, 8P/Tuttle, and 103P/Hartley 2, and even 19P/Borrely and 1P/Halley, were formed as or evolved to be contact binaries (or more generally, bi-lobate bodies) (Harmon et al., 2010; Rickman et al., 2015, Nesvorný et al., 2018), albeit from smaller components than the ostensible components of Ultima. Gravitational collapse of "pebble clouds" in the solar nebula is predicted to have naturally and efficiently formed binaries in the Kuiper belt, and with subequal mass ratios (Nesvorný et al., 2010). Some contact and near-contact binary bodies may form by rotational fission (e.g., Walsh et al., 2008), though this relies on a thermal YORP mechanism (effective for small asteroids but relatively ineffective in the Kuiper Belt and for larger bodies) and generally yields small secondary/primary mass ratios.

Close binaries can also form from wider binaries, and contact binaries from close binaries, through tidal dissipation of orbital energy, though angular momentum must be removed as well, such as by collisional diffusion, shedding material from the system or by modifying the spin states of the bodies (e.g., Porter and Grundy, 2012; Nesvorný et al., 2018). Three-body systems can evolve such that the third body is ejected, carrying off energy and angular momentum and allowing the remaining two to become more tightly bound. Even a single body, if elongated and rotating, can become bi-lobate as the result of a sub-catastrophic collision, and at impact velocities well under 1 km s[-1] (Jutzi and Benz, 2017; Schwartz et al., 2018). A virtue of being a true binary, as opposed to a contact binary or bilobate body, is that the mass of the system can be estimated (and thus the mean density of Ultima constrained from its total volume) from the size and period of their mutual orbit; determination of the system barycenter from *New Horizons* images would give the densities of the individual components.

Our current lack of information prevents us from knowing whether objects like Ultima entirely form from the low velocity accretion of small (1-mm to 1-m-scale) building blocks, variously referred to as spherules or pebbles in the astrophysical literature (e.g., Johansen & Lacerda, 2010; Johansen et al., 2015; Blum et al., 2017).  If this is the case, then it might contribute the appearance of a mantle at the best resolution *New Horizons* will image Ultima (several tens of meters).  Conversely, if Ultima's overall appearance is strongly faceted, this would imply it is a coherent fragment from disruption of a larger parent body, though it has also been argued that catastrophic disruption of a porous, low-strength body may also yield a body similar to 67P (Morbidelli & Rickman, 2015; Jutzi and Benz, 2017; Jutzi et al., 2017).

One of the more remarkable structural features of 67P was the pervasive layering at the surface discovered in *Rosetta* imaging (Massironi et al., 2015). Though layering had been seen or suspected in other comets, the level of detail provided by Rosetta high-resolution imaging was unprecedented. An approximately 650-m-thick stack of layers, individually between 50 and 200 m thick, envelops both lobes of 67P. The origin of these layers is enigmatic and not agreed upon, but is clearly a major clue to the assembly of the comet (cf. Massironi et al., 2015; Jutzi et al., 2017; Schwartz et al., 2018). The best LORRI images of Ultima (~35 m px[-1]) might just be able to detect layering of the thicknesses seen at 67P, and more easily if layer thicknesses scale with the size the body. Detection and characterization of such layers and other surface and structural features of Ultima is a major goal of the upcoming *New Horizons* encounter, and should advance our understanding of accretion and catastrophic disruption in the Kuiper Belt and thus our overall understanding of early Solar System evolution.

## 7. Moons and Rings

During the approach to Ultima, we will conduct a deep search for small satellites and rings with the LORRI camera. Small satellites have previously been detected around several KBOs (e.g.,

Brown et al., 2006; Stephens and Noll, 2006; Parker et al., 2016; Kiss et al., 2017), Centaurs (10199) Chariklo and (2060) Chiron, and dwarf planet Haumea have rings (Braga-Ribas et al., 2014; Ortiz et al., 2015, 2017), and the moderately-sized (47171) Lempo is a hierarchical triple system (Bennechi et al., 2010).

This search is not just of scientific interest; it is needed to assess any potential hazards to the spacecraft during the flyby. Based on the probable size of Ultima's Hill sphere, satellite orbits could be stable out to distances of ~100,000 km. Simulations of approach phase images indicate that a moon as small as 1 km could be detected, assuming a geometric albedo ~ 0.1. However, because orbital periods can be very long—decades—it will be difficult to determine the precise orbit of a moon unless it orbits within a few thousand kilometers of Ultima. Narrow rings could potentially be detected in the existing stellar occultation data (Young et al., 2017). Detecting an optically thin dust ring requires LORRI or MVIC. Simulations show that most dust is orbitally unstable near Ultima because solar radiation pressure, although weak, is still stronger than Ultima's gravity. However, Hamilton (2018) has shown that dust can be stable in a "sunflower" configuration, where the orbits are oriented perpendicular to the Sun. Solar radiation pressure keeps the dust in this configuration as Ultima orbits the Sun. Our studies using simulated ring images indicate that LORRI imaging on approach will be capable of detecting broad rings of width 1000 km to an I/F (reflected/incident solar radiation) of ~3e-8. This limit is deeper than observed at Pluto, due to the reduced stray light of the central body, and the use of longer LORRI exposures than were taken at Pluto (Lauer et al., 2018). Additional high-phase imaging with MVIC may be able to search to even lower optical depth limits.

## 8. Final Comments

Just as with our encounter with the Pluto system, The *New Horizons* encounter with Ultima will completely revolutionize our understanding of the small denizens of the Kuiper Belt. Ultima, being a cold classical object, is a particularly exciting opportunity for a close-up investigation of an object probably resident in the outer Solar System since the time of its formation. The story it should tell of solar system origins and the subsequent environment along the Solar System's edge since its formation may well place end member values on such topics as impact flux rates and volatile compositions. The encounter with Ultima fills a critical gap in our knowledge about KBOs in its size range: Pluto represents the largest KBO, which underwent substantial differentiation (Stern et al., 2018), while *Rosetta*'s close scrutiny of 67P/Churymov-Gerasimenko revealed the primitive nature of a small (~4-km wide) object from the Kuiper Belt (Taylor et al., 2017). Pluto's moon Charon and Saturn's Phoebe, which appears undifferentiated and quite possibly originated in the Kuiper Belt (Johnson and Lunine, 2005; Castillo-Rogez et al., 2012), represent medium-sized KBOs, while Ultima fills a critical gap in size between these objects and 67P.

Ultima's exploration —the furthest robotically-visited world in the Solar System—has important implications for better understanding the original context of cometary nuclei, the origin of small planets, the Solar System as a whole, the solar nebula, and extra-solar disks, as well as studying thermally primitive material from the planet formation era. This exploration will transform Kuiper Belt and Kuiper Belt object science from a purely astronomical regime, to a geological and geophysical regime, which radically changed paradigms when the same happened to asteroids and comets in past decades.

## Acknowledgements

We thank the many people, especially the *New Horizons* Science Team and Engineering Team, who have worked very hard in myriad ways to make the *New Horizons* mission and the plans for the encounter described here possible. Special thanks to the anonymous reviewer whose suggestions improved the content and clarity of the manuscript. This work was funded by NASA's *New Horizons* Project. All data from the Ultima Thule encounter, once losslessly received and calibrated, will "ultimately" be posted on the NASA Planetary Data System.

# Figures

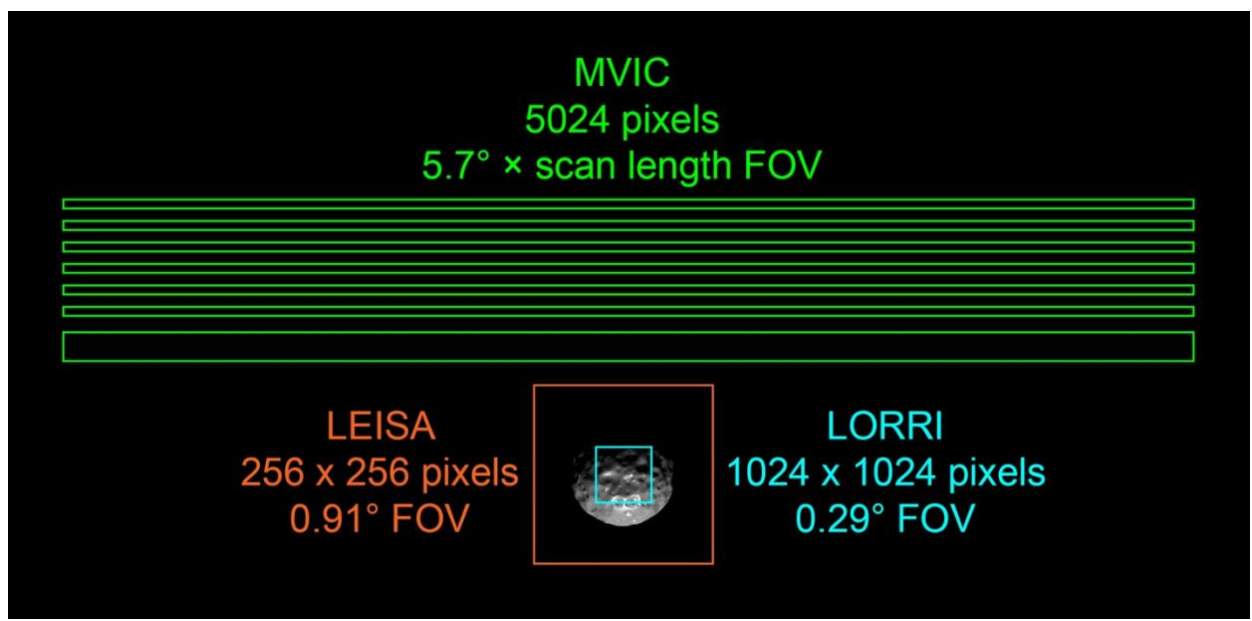

**Figure 1.** Shown here are the relative Fields of View (but not their actual boresights) of the principal remote sensing instruments on *New Horizons* at the time of closest approach (~05:33 UTC). The model representing Ultima is 30 km across. Best (highest resolution) encounter imaging will take place a few minutes before closest approach in order to optimize surface coverage and lighting (phase angle). Although all three instrument fields-of-view are shown here for reference, only MVIC and LORRI will operate during the final observation before closest approach.

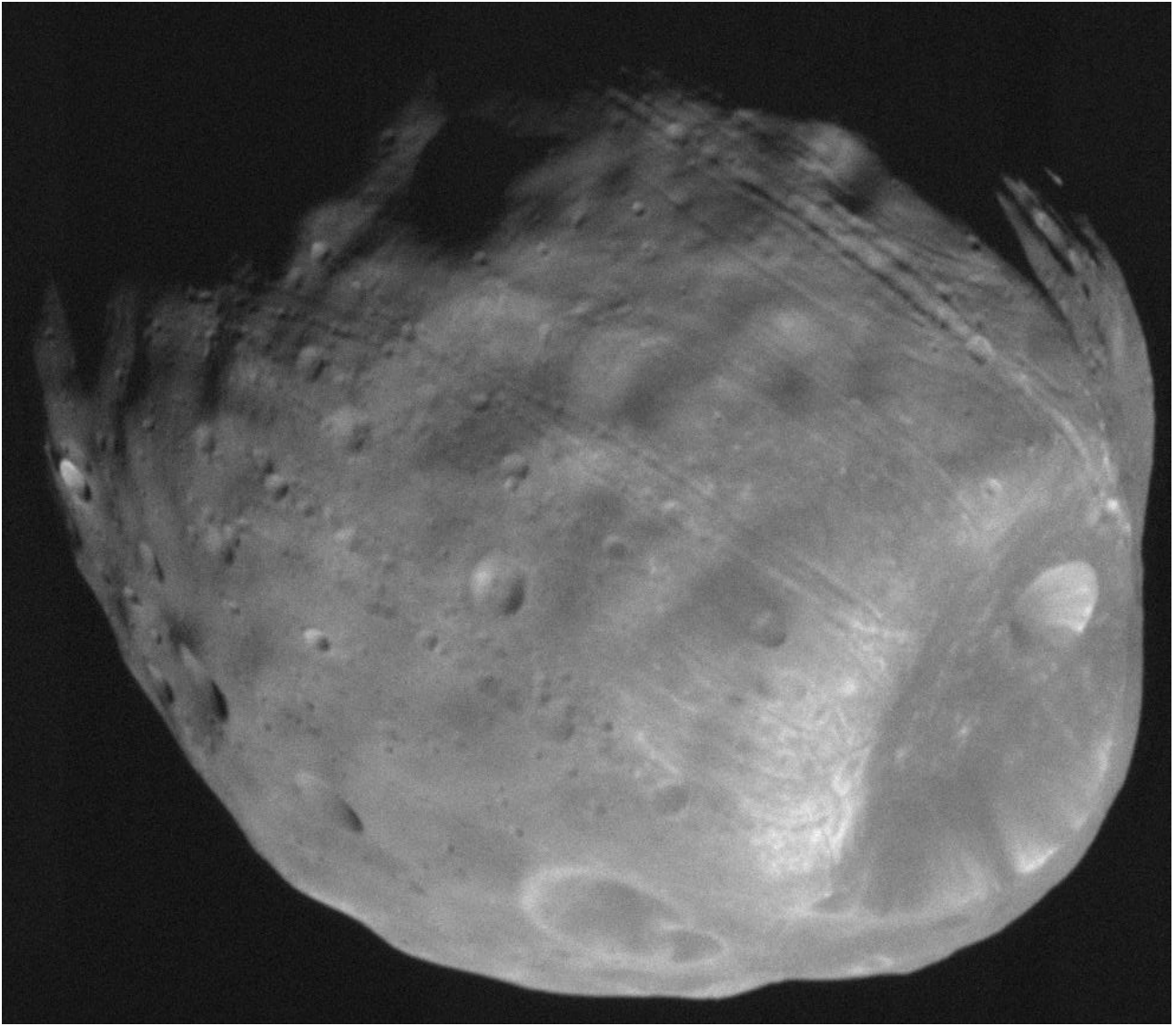

**Figure 2.** The best images of Ultima are planned to have an effective resolution between 42 and 78 m/pixel (as a consequence of smear and SNR effects) at a phase angle between 27 and 55°. What this might resemble can be demonstrated using this HiRISE image of Phobos (mean diameter 22.5 km), obtained at a phase angle of 26.4° and with resolution degraded to 78 m/pixel, with expected smear and SNR simulated.